\documentclass[twocolumn,showpacs,preprintnumbers,amsmath,amssymb,prl,superscriptaddress]{revtex4}

\usepackage{amsfonts}
\usepackage{graphicx}
\usepackage{dcolumn}
\usepackage{bm}

\begin{document}

\preprint{APS/123-QED}

\title{Twofold spontaneous symmetry breaking in a heavy fermion superconductor UPt$_3$}

\author{Y. Machida}
\author{A. Itoh}
\author{Y. So}
\author{K. Izawa}
\affiliation{Department of Physics, Tokyo Institute of Technology, Meguro 152-8551, Japan}
\author{Y. Haga}
\author{E. Yamamoto}
\affiliation{Advanced Science Research Center, Japan Atomic Energy Agency, Tokai 319-1195, Japan}
\author{N. Kimura}
\affiliation{Department of Physics, Tohoku University,
Sendai 980-8577, Japan}
\author{Y. Onuki}
\affiliation{Advanced Science Research Center, Japan Atomic Energy Agency, Tokai 319-1195, Japan}
\affiliation{Department of Physics, Osaka University, Toyonaka 560-0043, Japan}
\author{Y. Tsutsumi}
\affiliation{Department of Physics, Okayama University, Okayama 700-8530, Japan}
\author{K. Machida}
\affiliation{Department of Physics, Okayama University, Okayama 700-8530, Japan}

\date{\today}

\begin{abstract}
The field-orientation dependent thermal conductivity of the heavy-fermion superconductor UPt$_3$ 
was measured down to very low temperatures and under magnetic fields
throughout three distinct superconducting phases: A, B, and C phases.
In the C phase,
a striking twofold oscillation of the thermal conductivity within the basal plane is 
resolved reflecting the superconducting gap structure with a line of node along the $a$ axis.
Moreover, we find an abrupt vanishing of the oscillation across a transition to the B phase,
as a clear indication of a change of gap symmetries.
We also identify extra two line nodes below and above the equator in both B and C phases.
From these results together with the symmetry consideration, the gap function of UPt$_3$ is conclusively determined as a $E_{1u}$ representation
characterized by a combination of two line nodes at the tropics and point nodes at the poles.
\end{abstract}


\maketitle
Spontaneous symmetry breaking is one of the fundamental paradigms encompassing from
condensed matter physics to high energy physics, constituting the foundation of modern physics.
This paradigm is crucial sometimes because it can give a handle to discover
some unknown exotic ordered phase.
This is particularly true when broken symmetry is extremely low, that is,  the ``residual symmetry''
is so small, one may effectively and self-evidently narrows down possible
ordered phase to identify.

Understanding the unconventional superconductivity, in which electron pairs are formed without phonon, 
has been a challenge.
Part of the problem in uncovering the mechanism is that little is known about the pairing symmetry.
The heavy-fermion superconductor UPt$_3$ is one of the examples whose pairing symmetries are as yet to be clarified.
The most intriguing feature of this material is the  existence of a multiple phase diagram;
UPt$_3$ undergoes a double superconducting transition 
at the upper critical temperature $T_{c}^{+}$ $\sim$ 540 mK into the
A phase and at
the lower critical temperature $T_{c}^{-}$ $\sim$ 490 mK into
the  B phase~\cite{fisher}. 
In addition, the third (C) phase is stabilized at low temperatures under high magnetic fields~\cite{adenwalla}.
A crucial role of a weak antiferromagnetic order below $T_{\rm N}\sim$ 5 K for the phase multiplicity is
indicated by the pressure studies~\cite{hayden}.
Power law dependence of the thermodynamic and transport quantities reveal the
presence of nodes in the superconducting gap~\cite{shivaram,kohori,brison,suderow}. 
Moreover, a possibility of an odd-parity pairing is inferred from 
the nuclear magnetic resonance studies of the
Knight shift~\cite{tou}
and is supported theoretically~\cite{sauls}
by eliminating the singlet even parity scenario.

Extensive theoretical efforts have been devoted to explain these disparate experimental results~\cite{sauls,park,machida}. 
Among them, the $E_{2u}$ scenario with a line node in the basal plane and point nodes along the $c$ axis
has been regarded as one of the  promising candidates~\cite{joynt}.
Several experimental results, such as the anisotropy of the thermal 
conductivity~\cite{lussier} and the ultrasonic attenuation~\cite{ellman}
as well as the recent small-angle neutron scattering~\cite{huxley} and the Josephson tunnel junction~\cite{strand},
have been claimed to be compatible with this model.
On the other hand, there exist some controversies in explaining the following experiments;
1) the spontaneous internal field due to the broken time-reversal symmetry is most likely absent~\cite{dalmas},
2) the $d$-vector has two components in the B phase~\cite{tou},
3) a point where the three superconducting phases meet is a tetracritical point~\cite{adenwalla}.
Moreover,  to date no experimental evidence for the gap structure of each phase associated with the $E_{2u}$ model has been provided.
The pairing symmetry of UPt$_3$, therefore, remains unclear.

One of the most conclusive ways to identify the pairing symmetry is to elucidate the gap structure 
by the thermal conductivity measurements with rotating magnetic fields relative to the crystal axes
deep inside the superconducting state.
This technique has been successful to probe the nodal gap structure of several unconventional superconductors
by virtue of its directional nature and sensitivity to the delocalized quasiparticles~\cite{matsuda}.
In this paper, we present a decisive experiment of the angular dependence of the thermal conductivity of UPt$_3$ revealing the spontaneous rotation symmetry lowering, namely
the unusual gap structure 
with a lower rotational symmetry than the crystal structure.

High quality single crystal of UPt$_3$ with the high residual  resistivity ratio of 800 was grown by the Czochralski pulling method in a tetra-arc furnace~\cite{kimura1}.
We measured the thermal conductivity along the hexagonal $c$ axis (heat current $q\parallel$ $c$) 
on the sample with a rectangular shape (3 $\times$ 0.42 $\times$ 0.4 mm$^3$). 
To apply the magnetic fields with high accuracy 
relative to the crystal axes, we used a system with two superconducting
magnets generating the fields in two mutually orthogonal directions.
The magnets are installed in a Dewar seating on a mechanical rotating stage,
enabling the continuous rotation of the magnetic fields.

First, we begin with demonstrating that the thermal conductivity ($\kappa$) well probes the superconducting quasiparticle (QP) structures
from its temperature ($T$) and magnetic field ($H$) dependences.
From now on, the hexagonal [$\bar{1}$2$\bar{1}$0], [$\bar{1}$010], and [0001] axes are denoted as the $a$, $b$, and $c$ axes, respectively.
The inset of Fig.~1 shows the $T$ dependence of $\kappa(T)/T$ under zero field and 3 T along the $b$ axis.
With decreasing $T$, the zero-field $\kappa(T)/T$ shows a steep increase up to $\sim$ 0.3 K without apparent anomalies at $T_c^+$ and $T_c^-$.
On further cooling, $\kappa(T)/T$ considerably decreases due to a reduction of the QP densities,
and takes an extremely small value at the lowest $T \sim T_c^+/20$, consistent with the previous measurements~\cite{suderow}.
In the normal state (3T), $\kappa(T)/T$ appears to continuously 
increase down to the lowest $T$.
The dashed line denotes $\kappa(T)/T$ obtained from the normal-state resistivity $\rho(T)$
using the Wiedemann-Franz law, $\kappa(T)/T=L_0/\rho(T)$ ($L_0$: the Lorentz number).
Importantly, we confirm that $\kappa(T)/T$ is close to $L_0/\rho(T)$ at low temperature $T<$ 100 mK, indicating the dominant electronic contribution in the heat transport.
In this $T$-range, the $H$ dependence of the thermal conductivity $\kappa(H)/T$ at 55 mK shows a remarkable $H$-linear dependence 
at low fields for both $c$ and $b$ directions (the main panel of Fig.~1)
in contradiction to the field-insensitive behavior of fully gapped superconductors except in the vicinity of $H_{c2}$~\cite{lowell},
providing evidence for the nodal superconductivity in UPt$_3$.

In addition, we find distinct anomalies associated with a transition from the B to C phase at $H_{\rm BC}$ (open arrows).
The fact that the BC transition manifests by a sharp change of the slope implies 
a suppression of one of the degenerate order parameter components in the B phase.
This behavior can be more clearly resolved for the $b$ axis.
The determined $H_{\rm BC}$ together with $H_{c2}$ denoted by the solid arrows are summarized in Fig.~3(d) for $H$ $\parallel$ $b$.
We also note that a striking anisotropy is found in $\kappa(H)/T$ at 55 mK near $H_{c2}$:
$\kappa/T$ for $H\parallel$ $c$ shows a rapid increase just below $H_{c2}$, while the one for $H\parallel$ $b$
linearly increases up to $H_{c2}$, as similarly observed in Sr$_2$RuO$_4$~\cite{izawa}.
A search of the relevance of this behavior to the odd-parity superconductivity is a fascinating issue to be addressed.
\begin{figure}[t]
\begin{center}
\includegraphics[scale =0.45]{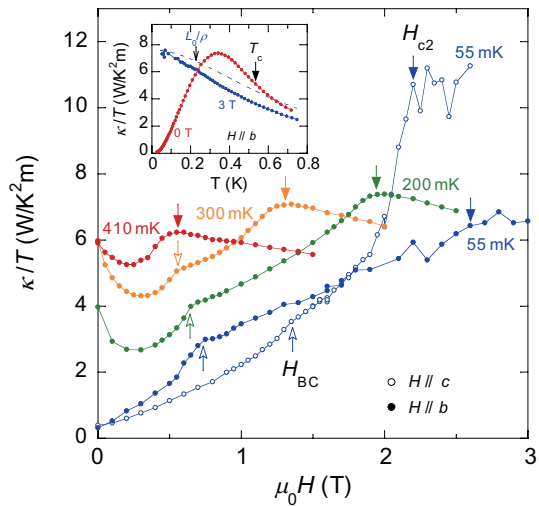}
\end{center}
\vspace{-0.5cm}
\caption{\label{fig.1} (color online).  Magnetic field dependence of the thermal conductivity $\kappa(H)/T$ along the $c$ and $b$ axes
at various temperatures. 
The open and closed arrows represent the B $\rightarrow$ C transitions $H_{\rm BC}$ and the upper critical fields $H_{c2}$, respectively. 
Inset: temperature dependence of $\kappa(T)/T$ under zero field and at 3 T for $H\parallel$  $b$.
The dashed line shows $\kappa/T=L_0/\rho$ ($L_0$: the Lorentz number) obtained from the normal-state resistivity $\rho$ 
using the Wiedemann-Franz law.}
\end{figure}

Next, to shed light on the nodal topology in the superconducting phases,
we concentrate on the angular dependence of $\kappa$.
The most significant effect on the thermal transport
for nodal superconductors in the mixed state comes from the Doppler shift
of the QP energy spectrum, $E(\bm{p}) \rightarrow E(\bm{p}) - \bm{v}_s \cdot \bm{p}$,
in the circulating supercurrent flow $\bm{v}_s$.
This effect
becomes important at such positions where the gap becomes smaller than the Doppler shift term
($\Delta< \bm{v}_s \cdot \bm{p}$).
The maximal magnitude of the
Doppler shift strongly depends on the
angle between the node direction and $H$,
giving rise to the oscillation of the density of states (DOS).
Consequently, $\kappa$ attains the maximum (minimum)
value when $H$ is directed to the antinodal (nodal) directions~\cite{vekhter}.
Figure~2 shows $\kappa(\phi)$ normalized by the normal state value $\kappa_n$ 
as a function of the azimuthal angle $\phi$ at 50 mK ($\sim T_c^+/10$)
at $|\mu_0H|$ = (a) 3.0 T, (b) 1.0 T, and (c) 0.5 T, respectively.
The data are taken in rotating $H$ after field cooling at $\phi$ = -70$^{\circ}$,
and $\kappa_n$ is measured at 50 mK above $H_{c2}$ for $H\parallel$ $b$.
In the normal state (3.0 T) and the B phase (0.5 T), we find no $\phi$-dependence within experimental error.

By contrast, what is remarkably is that $\kappa(\phi)$ exhibits a distinct twofold oscillation
with a minimum at $\phi$ = 0$^{\circ}$ in the C phase (1.0 T).
The open circles are obtained under field cooling condition at each angle.
The data obtained by different procedures of field cooling coincide well 
with each other, indicating negligibly small effect of the vortex pinning.
Strikingly, since the twofold symmetry is lower than the hexagonal crystal structure,
the in-plane anisotropy of the Fermi surface and $H_{c2}$~\cite{joynt} is immediately ruled out as the origin.
As shown by the solid lines, $\kappa(\phi)$ can be decomposed into two terms;
$\kappa(\phi) = \kappa_0 + \kappa_{2\phi}$, where $\kappa_0$ is a $\phi$-independent term and $\kappa_{2\phi}=C_{2\phi}\cos2\phi$ is a twofold component.
Figure~2(e) shows the amplitude of the twofold component $|C_{2\phi}/\kappa_n|$ as a function of $H/H_{c2}$,
where $H_{c2}$ = 2.6 T for $H\parallel$ $b$.
It can be clearly seen that $|C_{2\phi}/\kappa_n|$ suddenly appears to 
be finite $\sim 3 \%$ in the C phase,
implying a change of the gap symmetries across the BC transition that is of second order.
We note that $|C_{2\phi}/\kappa_n|$ obtained by rotating $H$
conically around the $c$ axis at fixed $\theta$ = 63$^{\circ}$ is same order of magnitude with the values at $\theta$ = 90$^{\circ}$
as denoted by an open circle in Fig.~2(e).
\begin{figure}[t]
\begin{center}
\includegraphics[scale =0.55]{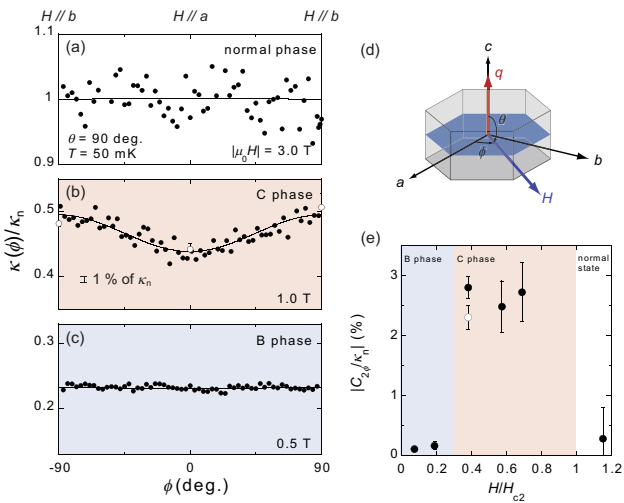}
\end{center}
\vspace{-0.5cm}
\caption{\label{fig.2} (color online). 
Angular variation of the thermal conductivity
$\kappa(\phi)$ normalized by $\kappa_n$ 
at 50 mK as a function of the azimuthal angle $\phi$ for $|\mu_0H|$ = (a) 3.0 T, (b) 1.0 T, and (c) 0.5 T, respectively. 
$\kappa(\phi)$ is measured with rotating $H$ within the $ab$ plane (the polar angle $\theta=90^\circ$)
as schematically shown in (d),
where $\phi$ and $\theta$ are measured from the $a$ and $c$ axes, respectively, and $q$ is injected along the $c$ axis.
The solid lines show the twofold component in $\kappa(\phi)/\kappa_n$.
The open circles represent $\kappa(\phi)/\kappa_n$ at 1 T obtained under the field cooling condition at every angle.
(e) Field variation of the twofold amplitude $|C_{2\phi}/\kappa_n|$ at 50 mK at $\theta=90^\circ$ (solid circles) and $63^\circ$ (open circle), respectively.}
\end{figure}

To further elucidate the gap symmetry,
we present the polar angle ($\theta$) dependence of $\kappa$ in Fig.~3,
showing
$\kappa(\theta)/\kappa_n$ measured by rotating $H$ within the $ac$ plane (green circles)
and the $bc$ plane (orange circles) at 50 mK at $|\mu_0H|$ = (a) 1.5 T, (b) 1.0 T, and (c) 0.5 T.
Here, $\kappa_n$ is measured at 50 mK above $H_{c2}$ for $H\parallel$ $c$.
The dominant twofold oscillation is found in all the  fields with maxima at $\theta$ = 90$^{\circ}$,
which could be attribute to, such as the Fermi surface and/or the gap topology or
the difference in transport with $H$ parallel to and normal to the heat current $q$.
Regardless of the origin,
the fact that $\kappa(\theta)/\kappa_n$ is maximized at $\theta$ = 90$^{\circ}$
excludes an artificial origin of the in-plane twofold oscillation in the C phase due to a misalignment
of $H$ relative to $q$.
We thus conclude that the in-plane twofold symmetry in the C phase is a consequence 
of the node.

In the B phase (0.5 T), the two different scanning procedures within the $ac$ and $bc$ planes well converge with each other,
consistent with the $\phi$-independence of $\kappa$.
In addition, we find extra two minima at $\theta$ = 20$^{\circ}$ and 160$^{\circ}$.
By plotting $\Delta\kappa(\theta)/\kappa_n\equiv(\kappa(\theta)-\kappa_0-\kappa_{2\theta})/\kappa_n$ vs $\theta$
after the subtraction of $\kappa_0$ and $\kappa_{2\theta}=C_{2\theta}\cos2\theta$, 
the minima become clearly visible 
at 35$^{\circ}$ and 155$^{\circ}$ (Fig.~3(c), inset).
This double-minimum structure is also found in the C phase (Fig.~3(a)).
We infer that these minima are derived from the two horizontal line nodes at the tropics as discussed below.
In contrast to the B phase, the two scanning results do not coincide in the C phase (Fig.~3(a)); 
the difference is diminished at the poles and maximized at $\theta$ = 90$^{\circ}$,
being consistent with the in-plane twofold symmetry.
Moreover, a significant appearance of the twofold symmetry across the BC transition can be seen at 1.0 T (Fig.~3(b)), in which
one experiences the BC (CB) transition twice by varying $\theta$
because of the anisotropy of $H_{\rm BC}$.
Indeed, the transitions occur at $\theta$ = 30$^{\circ}$ and 150$^{\circ}$ taking distinct kinks.
Remarkably, the difference between the two scanning procedures becomes finite upon entering the C phase,
providing the compelling evidence for the twofold symmetry of the gap structure in the C phase.
Moreover, the fact that $|C_{2\phi}/\kappa_n|$ takes same order of the magnitude at $\theta$ = 90$^{\circ}$ and 63$^{\circ}$
is in favor of a line node along the $a$ axis rather than the point nodes in the basal plane.
Notably, although a mechanism which fixes domains is a puzzle, the in-plane twofold symmetry of $\kappa(\phi)$ indicates 
a single superconducting domain.
\begin{figure}[t] 
\begin{center}
\includegraphics[scale =0.45]{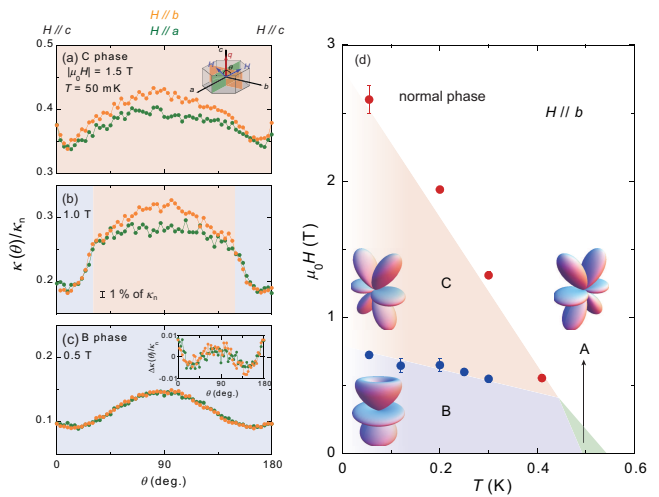}
\end{center}
\vspace{-0.5cm}
\caption{\label{fig.3} (color online). 
Angular variation of the thermal conductivity
$\kappa(\theta)$ normalized by $\kappa_n$
at 50 mK as a function of the polar angle $\theta$ for $|\mu_0H|$ = (a) 1.5 T,
(b) 1.0 T, and (c) 0.5 T, respectively.
The $\kappa(\theta)/\kappa_n$ curves measured by rotating $H$ (inset of (a)) within the $ac$ plane (green circles) and the $bc$ plane (orange circles)
are simultaneously plotted. 
Inset of (c): $\Delta\kappa(\theta)/\kappa_n \equiv (\kappa(\theta)-\kappa_0-\kappa_{2\theta})/\kappa_n$ vs $\theta$ plot
at 50 mK at 0.5 T, 
where $\kappa_0$ is a $\theta$-independent term and $\kappa_{2\theta}=C_{2\theta}\cos2\theta$ is a twofold component .
(d) The phase diagram of UPt$_3$ with the three
superconducting phases, labelled A, B, and C,
for $H\parallel b$. 
The red and blue circles represent 
$H_{\rm BC}$ and $H_{c2}$, respectively, deduced from the present measurements.
The schematic shapes of the gap symmetries for each phase are shown.}
\end{figure}

We discuss the order parameter symmetry of UPt$_3$ within the triplet category.
The present experiments indicate 
(i) the line node along the $a$ axis in the C phase,
(ii) the absence of in-plane gap anisotropy in the B phase, and
(iii) the two line nodes at the tropics in both B and C phases.
Taking into account all these results
and the $d$-vector configurations assigned by the Knight shift~\cite{tou},
the order parameter is unambiguously determined with a form of $(k_a\hat{b}+k_b\hat{c})
(5k_c^2-1)$ for the B phase,
where $\hat{b}$ and $\hat{c}$ are unit vectors of the hexagonal axes representing the directions of $d$-vectors.
This state belongs to two-dimensional $E_{1u}$ representation with the $f$-wave character,
the so-called planar state in triplet pairing
in the $D_{6h}$ hexagonal symmetry,
and to degenerate $E_u$ state for the recent claimed $D_{3d}$ trigonal symmetry~\cite{walko,note}. 
The gap structure consists of  the two horizontal line nodes at the tropics 
($k_c=\pm1/\sqrt{5}$, $\theta$ = 63$^{\circ}$ and 117$^{\circ}$) 
and the point nodes at the poles ($k_a=k_b=0$).
Note that although the locations of the horizontal line nodes estimated by assuming a spherical Fermi surface
do not agree with the observation ($\theta$ = 35$^{\circ}$ and 155$^{\circ}$), it could be changed by considering the realistic Fermi surface~\cite{joynt}.

By lifting the doubly degeneracy, the order parameter for the C phase
is given by $k_b\hat{c}(5k_c^2-1)$ for $H\parallel ab$ and $k_b\hat{a}(5k_c^2-1)$ 
for $H\parallel c$, respectively.
In the same manner, $k_a\hat{b}(5k_c^2-1)$ state is readily assigned  for the A phase.
The schematic shapes of the gap symmetries in the three phases are shown in Fig.~3(d).
We emphasize that this state is compatible not only with the hybrid gap state indicated by the several experiments~\cite{lussier,ellman},
in the sense that the line and point nodes simultaneously exist,
but also with some experimental results for which the $E_{2u}$ model~\cite{sauls} has failed to describe,
i.e., the absence of the internal magnetic field~\cite{dalmas}, the two-component $d$-vector for the B phase~\cite{tou}, 
and the tetracritical point in the phase diagram~\cite{adenwalla}.

To further strengthen our identification, in particular on the existence of the 
horizontal line nodes on the tropics, we calculate the angle-resolved DOS
by solving the Eilenberger equation~\cite{ichioka2} for several possible gap functions.
We compare here putative three gap functions in the C phase 
relative to the data in Fig.~4 where $\kappa(\theta)/\kappa_n$ and
the DOS differences along the vertical nodal and antinodal $\theta$-scannings are depicted.
The double peak structure characteristic in $E_{2u}$ and $E_{1g}$ whose 
origin comes from the horizontal node on the equator is not supported
by the data that are consistent with the present $E_{1u}$ with the horizontal nodes
on the tropics. In view of the Doppler shift idea mentioned above the
QPs in the horizontal node on the equator contribute more
when the field direction is away from $\theta=90^{\circ}$.
\begin{figure}[t] 
\begin{center}
\includegraphics[scale =0.9]{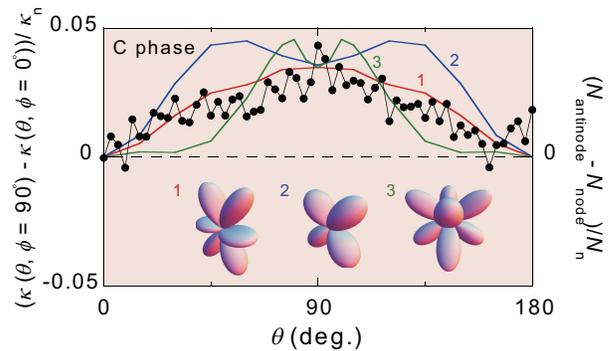}
\end{center}
\vspace{-0.5cm}
\caption{\label{fig.4} (color online). 
$\theta$-dependence of the thermal conductivity obtained by subtracting the
green data from the orange data in Fig.~3(a) (left axis) and the density of states  difference 
normalized at $\theta=90^{\circ}$ (right axis, arbitrary scale) along the vertical nodal
and antinodal scannings for three possible gap functions in the C phase:
1. The present $E_{1u}$ ($k_b(5k_c^2-1)$), 
2. $E_{1g}$ ($k_bk_c$),  3. $E_{2u}$ ($k_ak_bk_c$).
Those gap structures are sketched in the inset.}
\end{figure}

In summary, we find striking twofold oscillations in angle-resolved thermal conductivity
measurements at low temperatures
in a strongly correlated  heavy fermion superconductor UPt$_3$.
This spontaneous symmetry lowering, which is the lowest possible
rotational symmetry breaking in hexagonal crystal fortuitously and effectively 
narrows down the possible symmetry classes and leads us to uniquely identify
the pairing symmetry for each phase in the multiple phase diagram.
We conclude that the realized pairing function is $E_{1u}$ with the $f$-wave
character, i.e., the so-called planar state in the triplet pairing. 
This state is analogous to the B phase in superfluid $^3$He, and 
obviously bears the Majorana zero mode at a surface~\cite{chung,tsutsumi},
namely a topological superconductor that is quite rare to find.
Thus it is worth exploring further to  understand 
this interesting material as a new platform for topological physics.

We acknowledge insightful discussions with T. Ohmi, M. Ozaki, M. Ichioka,
H. Kusunose, and K. Ueda.
This work is partially supported by
grants-in-aid from the Japan Society for the Promotion of Science;
by grants-in-aid for Scientific Research on
Innovative Areas ``Heavy Electrons" (20102006) from
the Ministry of Education, Culture, Sports, Science, and Technology (MEXT), Japan;
and by Global COE Program from the MEXT through the Nanoscience and
Quantum Physics Project of the Tokyo Institute of Technology.

\bibliography{UPt3}

\begin{thebibliography}{27}
\bibitem[{not()}]{note}
\bibinfo{note}{The previously proposed $E_{1u}$ with the $p$-wave character
  such as $k_a+ik_b$ simply inappropriate because it yields no line node in the
  B phase.}
\end{thebibliography}
\end{document}